# Ge-Ge$_{0.92}$Sn$_{0.08}$ core-shell single nanowire infrared photodetector with superior characteristics for on-chip optical communication

Sudarshan Singh[1], Subhrajit Mukherjee[1,2], Samik Mukherjee[3,¥], Simone Assali[3], Lu Luo[3], Samaresh Das[4], Oussama Moutanabbir[3], and Samit K Ray[1]

## AFFILIATIONS

[1]Department of Physics, Indian Institute of Technology Kharagpur, Kharagpur, West Bengal 721302, India

[2]Presently at Faculty of Materials Science and Engineering, Technion-Israel Institute of Technology, Haifa, 3203003, Israel

[3]Department of Engineering Physics, École Polytechnique de Montréal, C. P. 6079, Succ. Centre-Ville, Montreal, Québec H3C 3A7, Canada

[¥]Present affiliation: Leibniz Institute for Solid State and Materials Research, Helmholtzstraße. 20, 01069 Dresden, Germany

[4]Centre for Applied Research in Electronics, Indian Institute of Technology Delhi, New Delhi 110016, India

Email: physkr@phy.iitkgp.ac.in

## ABSTRACT

Recent development on Ge$_{1-x}$Sn$_x$ nanowires with high Sn content, beyond its solid solubility limit, make them attractive for all group-IV Si-integrated infrared photonics at nanoscale. Herein, we report a chemical vapour deposition-grown high Sn-content Ge-Ge$_{0.92}$Sn$_{0.08}$ core-shell based single nanowire photodetector operating at the optical communication wavelength of 1.55 μm. The atomic concentration of Sn in nanowires has been studied using X-ray photoelectron and Raman spectroscopy data. A metal-semiconductor-metal based single nanowire photodetector, fabricated via electron beam lithography process, exhibits significant room-temperature photoresponse even at zero bias. In addition to the high-crystalline quality and identical shell composition of the nanowire, the efficient collection of photogenerated carriers under an external electric field result in the superior responsivity and photoconductive gain as high as ~70.8 A/W and ~57, respectively at an applied bias of -1.0 V. The extra-ordinary performance of the fabricated photodetector demonstrates the potential of GeSn nanowires for future Si CMOS compatible on-chip optical communication device applications.

Being compatible to complementary metal-oxide-semiconductor (CMOS) semiconductor technology, germanium−tin (GeSn) alloys have the potential to substantially increase the performance of optoelectronic devices functioning in the short-wave infrared (SWIR) range (0.9 - 2.5 μm) for various photonics applications including telecommunication and sensing[1–4]. However, the indirect-to-direct bandgap crossover of unstrained GeSn alloy occurs at a Sn content in the range of ~6.5 – 11 at. %,[2,5,6] which is much larger compared to the equilibrium solubility of Sn in Ge (<1.0 at. %)[5,7]. Therefore, when grown on Ge virtual substrates, the Sn-rich GeSn epilayer develops a high degree of strain due to the lattice mismatch resulting in the formation of extended defects (misfit dislocations) at the interface, which act as Shockley-Read-Hall recombination centres degrading the radiative lifetime of minority carriers and provide leakage currents in the device[8,9]. Recently, enhanced strain relaxation with high crystallinity has been realized in metal-assisted vapour-liquid-solid (VLS) grown Ge-Ge$_{1-x}$Sn$_x$ core-shell nanowires (NWs) at low temperatures[8,10–12], which inhibits the development of structural defects by shifting some strain into the Ge-core[13] leading to lower leakage current in photodetector[14]. In addition, the tensile misfit strain in the Ge-core may assist in lowering the Γ- to L-conduction valley energy that could constitute a dual direct bandgap material to improve the device performance[15].

Owing to the inherent high surface-to-volume ratio and smaller channel dimensions, a single nanowire based photodetector leads to very high photosensitivity, faster speed, and low power consumption[16–18]. In particular, Ge-Ge$_{1-x}$Sn$_x$ core-shell NWs demonstrated remarkable optical emission properties covering near-infrared (NIR) to SWIR spectral range due to its lower energy gap and high crystalline quality[10–12]. On account of superior optical emissivity and enhanced absorption due to narrow bandgap, Ge-Ge$_{1-x}$Sn$_x$ core-shell NWs appear potentially attractive as nanoscale photodetector for the optical communication and sensing devices. Recently, Yang et al.[19] reported the photodetection characteristics of molecular-beam-epitaxy (MBE) grown GeSn/Ge dual-nanowire, with a low dark current of 40 nA at 10 mV and an extended cut-off wavelength of 2.2 µm. However, the responsivity of the device at 1.55 µm is relatively lower compared to other GeSn thin film- based devices[20,21].

Here, we present the room-temperature photodetection characteristics of chemical vapour deposition (CVD) grown undoped single Ge-Ge$_{0.92}$Sn$_{0.08}$ core-shell NW at a wavelength of 1.55 µm. Ge-Ge$_{0.92}$Sn$_{0.08}$ core-shell NWs were grown on Si (111) wafers via catalyst-assisted growth paradigms using a CVD system. To accomplish the core-shell NW geometry, firstly, arrays of Ge NWs were grown at 330 ℃ with germane (GeH$_4$), H$_2$ and a 2-nm thin Au layer as the precursor, the carrier gas, and the catalyst, respectively. Next, the GeSn shell was grown at 310 ℃ for 20 minutes using GeH$_4$ and tin-tetrachloride (SnCl$_4$), where the Ge/Sn ratio in the gas phase was ~120. The detailed results on the growth kinetics of the Ge-Ge$_{0.92}$Sn$_{0.08}$ core-shell NWs has been reported previously and can be found elsewhere[13]. The typical field-emission scanning electron microscopy (FESEM) images in Fig. 1(a) and 1(b) displays the cross-sectional and planar view of Ge-Ge$_{0.92}$Sn$_{0.08}$ core-shell NWs arrays, respectively. The vertically standing NWs do not reveal any bending, indicating the uniform distribution of strain in all directions along the radius at the interface of Ge-core and GeSn-shell[22]. The top and bottom diameters of the NW are found to be around 200 nm and 120 nm, respectively, with an average length of 2-3 µm (supplementary material, Fig. S1). An enhanced tip diameter with tapered top section of the NWs may be attributed to enhanced precursor decomposition induced via catalytic effect of the Au-Sn droplet[8,22]. A typical low magnification transmission electron microscopy (TEM) image of a few Ge-Ge$_{0.92}$Sn$_{0.08}$ core-shell NWs is depicted in Fig. 1(c). The micrograph reveals a uniform surface without any observed kinks or cluster suggesting nearly identical distribution of Sn into the entire shell of the nanowire without any segregation. A hemi-spherical metal seed at the tip of the single Ge-Ge$_{0.92}$Sn$_{0.08}$ core-shell NW can be clearly seen in a magnified TEM image (Fig. 1(d)), which further validates the catalyst driven VLS growth of NWs.

For quantitative analysis, X-ray photoelectron spectroscopy (XPS) was employed to examine the compositional ratio and chemical bonding of Ge-Ge$_{0.92}$Sn$_{0.08}$ core-shell NWs using Al−Kα radiation of energy 1486.6 eV. Prior to XPS analysis, the samples were subjected to Ar$^+$ sputter etching in an ultra-high vacuum (UHV) XPS chamber for a few seconds to eliminate the surface contamination and native oxide. Figure 2(a) depicts the Ge-*3d* and Sn-*3d* core level photoelectron spectra of Ge-Ge$_{0.92}$Sn$_{0.08}$ core-shell NWs. Observed XPS peaks arise from the GeSn shell due to the low inelastic mean free path of photoelectrons. A strong peak centered at ~29.0 eV corresponds to the binding energy of Ge-*3d* electron with no signatures of Ge oxides, indicating the presence of elemental Ge without any oxidation. Two distinct peaks centered at ~484.7 eV and ~493.1 eV are ascribed to the binding energy of Sn $3d_{5/2}$ and $3d_{3/2}$ electrons, respectively. The atomic concentration of elemental Sn in the GeSn shell has been estimated using the following relation[23],

$$C_{Sn}(\%) = \frac{(I_{Sn}/S_{Sn})}{(I_{Sn}/S_{Sn}) + (I_{Ge}/S_{Ge})} \times 100 \quad (1)$$

where I represents the intensity and S is the atomic sensitivity factor of the corresponding elements. The atomic concentration of Sn from the above equation has been estimated to be around ~8.0 %. Furthermore, to probe the local chemical bonding more precisely, the Raman spectra of Ge-Ge$_{0.92}$Sn$_{0.08}$ core-shell NWs have been recorded at room temperature using a 514.5 nm argon-ion laser source with calibrated Si-Si Raman peak at 520 cm$^{-1}$. Figure 2(b) presents the Raman spectrum of Ge-Ge$_{0.92}$Sn$_{0.08}$ core-shell NWs on Si substrate, and a spectrum from bulk Ge for reference. The Raman peak attributed to Ge-Ge LO phonon mode appearing at a lower wavenumber of ~295.8 cm$^{-1}$ in Ge-Ge$_{0.92}$Sn$_{0.08}$ core-shell NWs in comparison to that of bulk Ge. Using the reported model to decouple the strain and composition from the Raman spectra of epitaxial GeSn thin-films[24], the red-shift in the Raman line of the NWs corresponds to an uniaxial strain of -2.6 % and a Sn composition of ~8.3 % (supplementary material, Fig. S2), the latter being in agreement with the XPS data.

A metal-semiconductor-metal (MSM) device with the above core-shell structure, featuring easier integration without any requirement of doping for junction formation, was fabricated to realize a single NW-based photodetector device. For device fabrication, the NWs were separated from the substrate via sonication in ethanol, which was dispersed onto a pre-patterned oxidized Si substrate ($SiO_2$ thickness ~300 nm). An isolated NW with a longer length (~3.3 μm) was identified and connected to the pre-patterned electrodes using standard electron-beam-lithography (EBL) process followed by the deposition of Cr/Au as electrodes. Figure 3(a) depicts the schematic representation of a typical MSM photodetector device fabricated using a single Ge-$Ge_{0.92}Sn_{0.08}$ core-shell NW. A FESEM image of the fabricated device is shown in Fig. 3(b). Figure 3(c) displays the room temperature semi-logarithmic current-voltage (I-V) characteristics of the fabricated single NW device, measured under dark condition. The device depicts a typical MSM transport behaviour attributed to the formation of two back-to-back Schottky barriers at the metal-semiconductor (M-S) interfaces and exhibit a fairly low dark current in the order of tens of nA at $\pm$ 1.0 V. The photoresponse characteristics of the Ge-$Ge_{0.92}Sn_{0.08}$ core-shell single NW photodetector has been investigated by modulating the current as a function of time under pulsed laser excitation at a wavelength of 1.55 μm at room temperature. A substantial change in the device current can be observed under irradiation of light without applying any external bias, as shown in Fig. 3(d). At zero bias, the device reveals an extremely low dark current in the range of 1-2 pA (OFF-state) which instantaneously switches to a higher conducting state exhibiting a current value of ~25 pA (ON-state) under light illumination. The generation of photoinduced electron-holes pairs enhances the current through the device under irradiation of light. Such stable and repeatable periodic photoresponse at constant power and zero bias are consistently observed for more than 10 successive cycles. The observed zero-bias photoresponse in the Ge-Ge$_{0.92}$Sn$_{0.08}$ core-shell single NW photodetector is attributed to the presence of a built-in electric field originating from the existence of inhomogeneous interface states[25,26] and quasi-symmetrical area contacts[27,28] at the M-S interface. The built-in field at the M-S interface and smaller channel length of the fabricated NW device lead to an efficient separation and collection of photogenerated charge carriers at both the electrodes, which results in the optical response of the fabricated device even without any external bias.

The effect of incident power on the photocurrent of single NW photodetector has been studied by monitoring the transient photoresponse under modulated laser power intensity at a wavelength of 1.55 μm. Figure 4(a) illustrates the dynamic photoresponse characteristics of the fabricated Ge-Ge$_{0.92}$Sn$_{0.08}$ core-shell single NW photodetector as a function of excitation intensity at zero applied bias. The photocurrent enhances almost linearly with increasing incident power intensity and found to be nearly 10 times higher under an incident light intensity of 6.37 mW/cm$^2$, as compared to that of 0.95 mW/cm$^2$. For quantitative analysis, the measured photocurrent ($I_{ph}$) at different excitation power intensity (P) has been plotted and fitted to the power-law relation $I_{ph} \propto P^{\gamma}$, (Fig. 4(b)), where 'γ' determines the effect of incident light on the photocurrent. The value of exponent (γ = 0.98) is found to be slightly below unity, indicating that the photocurrent in Ge-Ge$_{0.92}$Sn$_{0.08}$ core-shell single NW photodetector is partially governed by the complex processes of carrier generation and recombination[19,25,29].

The key figure-of-merits of the photodetector, such as the responsivity ($R_\lambda$) and photoconductive gain ($G_{ph}$) at a wavelength of 1.55 μm, have been estimated using the following relations[30], respectively,

$$R_\lambda = \frac{\Delta I}{P_{opt}(\lambda)} \quad (2)$$

$$G_{ph} = R_\lambda \times \frac{hc}{e\lambda} \quad (3)$$

where $\Delta I$ and $P_{opt}$ are the photocurrent and incident power on the effective area of device, respectively. Other symbols in equation (2) and (3) have their usual meaning. The single NW was uniformly illuminated with a beam diameter of ~1.0 mm. The effective area of the NW exposed for light absorption is taken as $(2\pi r \times L)/2$, where r and L are the average radius (r = 80 nm) and length of the NW (L = 1.5 μm) between the electrodes, respectively. The illumination power on the device has been calibrated using an InGaAs detector (Model no. DET10D/M Thor labs.) Figure 4(c) depicts the room temperature power-dependent photo-responsivity of the fabricated single NW photodetector measured at zero bias. At a low power, the calculated responsivity value is found to be ~0.5 A/W, which increases to ~1.1 A/W as the incident power increases and remains almost constant at higher power. A nearly monotonic increment in the photocurrent as a function of incident light intensity ($\gamma = 0.98$) results in almost uniform photosensitivity of the single NW device at zero bias. However, a slight decrement in the responsivity at a higher incident power may be attributed to the surface recombination owing to the presence of finite surface states due to large surface-to-volume ratio in nanowires[29]. Indeed, the nanowire typically possesses large density of surface states, which originate due to dangling bonds, anti-sites and vacancies on the surface, resulting in pinning the Fermi-level. Consequently, the energy band bending leads to the accumulation of photogenerated electrons underneath the surface[31,32]. At a higher optical power, the decrement of responsivity is attributed to two factors: firstly the saturation of carrier density at the traps and secondly, an enhanced electron concentration in the NW center under high light intensity lowering the potential barrier, both of which contribute in reducing the lifetime of carriers. A combination of the above two processes may lead to the decrease in responsivity, in agreement with the observations of power dependent responsivity for different NW based photodetectors[32–34]. This is to be noted that our zero-bias responsivity value is two orders of magnitude higher than that reported recently in GeSn/Ge dual-nanowire based photodetector at 1.0 mV[19]. The extracted photoconductive gain of the fabricated detector is found to be as high as 0.95 at zero bias.

A remarkable increment in the responsivity of the fabricated Ge-Ge$_{0.92}$Sn$_{0.08}$ core-shell single NW photodetector is observed on application of an external bias. The bias-dependent responsivity has been measured by recording the dynamic photoresponse under dark and illumination with constant incident optical power. For a small applied bias of 0.1 V, the responsivity value increases more than 25 times than that observed at zero bias for the incident power of 24 pW and is found to be ~25.8 A/W. With increasing bias, the depletion region width at the M-S junction expands which enhances the density and collection efficiency of photogenerated carriers between the two electrodes, resulting in an enhancement of photocurrent[16,29]. The bias-dependent responsivity and photoconductive gain of the fabricated single NW photodetector at a wavelength of 1.55 μm are depicted in Fig. 4(d). The responsivity of the fabricated Ge-Ge$_{0.92}$Sn$_{0.08}$ core-shell single NW photodetector is observed to be as high as ~70.8 A/W at -1.0 V. The measured responsivity is found to be much higher than most of the reported single NW photodetector devices based on Ge and other different semiconductors and approximately four order higher than that reported in GeSn/Ge dual-nanowire photodetector at the communication wavelength of 1.55 μm (supplementary material, performance comparison Table SI). On the other hand, the photoconductive gain of the single NW photodetector device is found to be 57 at a bias of -1.0 V. The reduced carrier transit time under the presence of an external electric field[16] and smaller effective channel length of the device[18] result in a high photoconductive gain (more than unity), and hence the device exhibits a very high responsivity. Furthermore, the specific detectivity is also an important figure-of-merit of a photodetector, which usually estimates the smallest detectable optical signal and can be defined as[35,36]

$$D^*(\lambda) = \frac{(A\Delta f)^{1/2} R_\lambda}{i_n} \quad (4)$$

where A is the NW cross-sectional area of the Ge-Ge$_{0.92}$Sn$_{0.08}$ core-shell NW, Δf is the electrical bandwidth in Hz and $i_n$ is the noise current. Taking into consideration the shot noise from dark current and thermal noise current limited detectivity, the noise current has been evaluated using equation S3. The detectivity value of the fabricated Ge-Ge$_{0.92}$Sn$_{0.08}$ core-shell single NW photodetector is estimated to be ~8.4×10$^9$ Jones at -1.0 V for 1.55 μm. To the best of our knowledge, the device performance of the Ge-Ge$_{0.92}$Sn$_{0.08}$ core-shell single NW is found to be superior to most of the GeSn thin film based photodetectors reported till date (supplementary material, performance comparison Table SII) at the optical communication wavelength of 1.55 μm. The high surface-to-volume ratio and lower density of structural defects due to the core-shell geometry result in the superior performance of the single nanowire photodetector device. Our results demonstrate the potential of single GeSn core-shell NW device for future nanoscale photodetectors working at an optical communication wavelength of 1.55 μm.

In summary, this study demonstrates superior photodetection characteristics of a silicon-compatible single nanowire photodetector based on CVD-grown Ge-Ge$_{0.92}$Sn$_{0.08}$ core-shell NWs with a Sn content as high as ~8.0 %. Micro-structural characterizations reveal a cluster- free uniform distribution of Sn in the entire Ge-Ge$_{0.92}$Sn$_{0.08}$ core-shell nanowires. The simple MSM based Ge-Ge$_{0.92}$Sn$_{0.08}$ core-shell single NW photodetector exhibits a relatively low dark current (tens of nA) at an applied bias of ±1.0 V. The fabricated single NW device display a remarkably high room temperature responsivity of ~ 1.1 A/W at zero bias and

~70.8 A/W at -1V for the optical-communication wavelength of 1.55 µm, which are much higher in comparison to the previously reported GeSn based photodetectors. A high surface-to-volume ratio, low structural defects and smaller carrier transit path in the NW device result in very high photosensitivity and gain (~57). We believe that the Ge-Ge$_{0.92}$Sn$_{0.08}$ core-shell NW device demonstrated in this work paves a new avenue of group IV semiconductor-based nanoscale infrared photodetectors for optoelectronic integrated systems.

See the supplementary material for the FESEM image of Ge-Ge$_{0.92}$Sn$_{0.08}$ core-shell NW, NWs length histogram, decoupling of Sn composition and strain, noise current expression, and performance comparison tables of the fabricated single NW photodetector.

We would like to acknowledge partial financial support from MeitY and Department of Science and Technology (DST) (Government of India, Ministry of Science and Technology) supported NNetRA "SWI" Project. O.M. acknowledges support from NSERC Canada (Discovery, SPG, and CRD Grants), Canada Research Chairs, Canada Foundation for Innovation, Mitacs, PRIMA Québec, and Defence Canada (Innovation for Defence Excellence and Security, IDEaS). L.L acknowledges support from China Scholarship Council (CSC). SS and SKR would like to acknowledge Dr. Veerendra Dhyani for his support in device fabrication.

**AUTHOR DECLARATIONS**

**Conflict of Interest**

The authors declare no competing financial interest.

**DATA AVAILABILITY**

The data that support the findings of this study are available from the corresponding author upon reasonable request.

**REFERENCES**

[1] R. Soref, D. Buca, and S.-Q. Yu, Opt. Photonics News **27**, 32 (2016).

[2] S. Wirths, R. Geiger, N. Von Den Driesch, G. Mussler, T. Stoica, S. Mantl, Z. Ikonic, M. Luysberg, S. Chiussi, J.M. Hartmann, H. Sigg, J. Faist, D. Buca, and D. Grützmacher, Nat. Photonics **9**, 88 (2015).

[3] D. Zhang, C. Xue, B. Cheng, S. Su, Z. Liu, X. Zhang, G. Zhang, C. Li, and Q. Wang, Appl. Phys. Lett. **102**, (2013).

[4] O. Moutanabbir, S. Assali, X. Gong, E. O'Reilly, C.A. Broderick, B. Marzban, J. Witzens, W. Du, S.Q. Yu, A. Chelnokov, D. Buca, and D. Nam, Appl. Phys. Lett. **118**, (2021).

[5] S. Gupta, B. Magyari-Köpe, Y. Nishi, and K.C. Saraswat, J. Appl. Phys. **113**, 073707 (2013).

[6] T.R. Harris, M.-Y. Ryu, Y.K. Yeo, B. Wang, C.L. Senaratne, and J. Kouvetakis, J. Appl. Phys. **120**, 085706 (2016).

[7] S. Wirths, D. Buca, and S. Mantl, Prog. Cryst. Growth Charact. Mater. **62**, 1 (2016).

[8] S. Assali, A. Dijkstra, A. Li, S. Koelling, M.A. Verheijen, L. Gagliano, N. Von Den Driesch, D. Buca, P.M. Koenraad, J.E.M. Haverkort, and E.P.A.M. Bakkers, Nano Lett. **17**, 1538 (2017).

[9] K.C. Lee, M.X. Lin, H. Li, H.H. Cheng, G. Sun, R. Soref, J.R. Hendrickson, K.M. Hung, P. Scajev, and A. Medvids, Appl. Phys. Lett. **117**, (2020).

[10] S. Biswas, J. Doherty, D. Saladukha, Q. Ramasse, D. Majumdar, M. Upmanyu, A. Singha, T. Ochalski, M.A. Morris, and J.D. Holmes, Nat. Commun. **7**, 11405 (2016).

[11] A.C. Meng, C.S. Fenrich, M.R. Braun, J.P. McVittie, A.F. Marshall, J.S. Harris, and P.C. McIntyre, Nano Lett. **16**, 7521 (2016).

[12] M.S. Seifner, A. Dijkstra, J. Bernardi, A. Steiger-Thirsfeld, M. Sistani, A. Lugstein, J.E.M. Haverkort, and S. Barth, ACS Nano **13**, 8047 (2019).

[13] S. Assali, R. Bergamaschini, E. Scalise, M.A. Verheijen, M. Albani, A. Dijkstra, A. Li, S. Koelling, E.P.A.M. Bakkers, F. Montalenti, and L. Miglio, ACS Nano **14**, 2445 (2020).

[14] S. Singh, A.K. Katiyar, A. Sarkar, P.K. Shihabudeen, A.R. Chaudhuri, D.K. Goswami, and S.K. Ray, Nanotechnology **31**, 115206 (2020).

[15] A.C. Meng, M.R. Braun, Y. Wang, S. Peng, W. Tan, J.Z. Lentz, M. Xue, A. Pakzad, A.F. Marshall, J.S. Harris, W. Cai, and P.C. McIntyre, Mater. Today **40**, 101 (2020).

[16] C. Soci, A. Zhang, X.Y. Bao, H. Kim, Y. Lo, and D. Wang, J. Nanosci. Nanotechnol. **10**, 1430 (2010).

[17] S.K. Ray, A.K. Katiyar, and A.K. Raychaudhuri, Nanotechnology **28**, 092001 (2017).

[18] S. Mukherjee, K. Das, S. Das, and S.K. Ray, ACS Photonics **5**, 4170 (2018).

[19] Y. Yang, X. Wang, C. Wang, Y. Song, M. Zhang, Z. Xue, S. Wang, Z. Zhu, G. Liu, P. Li, L. Dong, Y. Mei, P.K. Chu, W. Hu, J. Wang, and Z. Di, Nano Lett. **20**, 3872 (2020).

[20] H. Tran, T. Pham, J. Margetis, Y. Zhou, W. Dou, P.C. Grant, J.M. Grant, S. Al-Kabi, G. Sun, R.A. Soref, J. Tolle, Y.-H. Zhang, W. Du, B. Li, M. Mortazavi, and S.-Q. Yu, ACS Photonics **6**, 2807

(2019).

[21] M.R.M. Atalla, S. Assali, A. Attiaoui, C. Lemieux- Leduc, A. Kumar, S. Abdi, and O. Moutanabbir, Adv. Funct. Mater. **31**, 2006329 (2021).

[22] S. Assali, M. Albani, R. Bergamaschini, M.A. Verheijen, A. Li, S. Kölling, L. Gagliano, E.P.A.M. Bakkers, and L. Miglio, Appl. Phys. Lett. **115**, 113102 (2019).

[23] R. Bar, A.K. Katiyar, R. Aluguri, and S.K. Ray, Nanotechnology **28**, (2017).

[24] Bouthillier, S. Assali, J. Nicolas, and O. Moutanabbir, Semicond. Sci. Technol. **35**, (2020).

[25] K. Das, S. Mukherjee, S. Manna, S.K. Ray, and A.K. Raychaudhuri, Nanoscale **6**, 11232 (2014).

[26] S. Sett, S. Sengupta, N. Ganesh, K.S. Narayan, and A.K. Raychaudhuri, Nanotechnology **29**, (2018).

[27] C. Zhou, S. Raju, B. Li, M. Chan, Y. Chai, and C.Y. Yang, Adv. Funct. Mater. **28**, 1 (2018).

[28] L. Dong, J. Yu, R. Jia, J. Hu, Y. Zhang, and J. Sun, Opt. Mater. Express **9**, 1191 (2019).

[29] X. Yan, B. Li, Y. Wu, X. Zhang, and X. Ren, Appl. Phys. Lett. **109**, 053109 (2016).

[30] J.W. John, V. Dhyani, S. Singh, A. Jakhar, A. Sarkar, S. Das, and S.K. Ray, Nanotechnology **32**, 315205 (2021).

[31] D. Wang, Y.L. Chang, Q. Wang, J. Cao, D.B. Farmer, R.G. Gordon, and H. Dai, J. Am. Chem. Soc. **126**, 11602 (2004).

[32] J. Li, X. Yan, F. Sun, X. Zhang, and X. Ren, Appl. Phys. Lett. **107**, 263103 (2015).

[33] X. Yan, B. Li, Y. Wu, X. Zhang, and X. Ren, Appl. Phys. Lett. **109**, (2016).

[34] A. Zhang, H. Kim, J. Cheng, and Y.-H. Lo, Nano Lett. **10**, 2117 (2010).

[35] X. Gong, M. Tong, Y. Xia, W. Cai, J.S. Moon, Y. Cao, G. Yu, C.-L. Shieh, B. Nilsson, and A.J. Heeger, Science (80-. ). **325**, 1665 (2009).

[36] X. Liu, L. Gu, Q. Zhang, J. Wu, Y. Long, and Z. Fan, Nat. Commun. **5**, 1 (2014).

**Figures:**

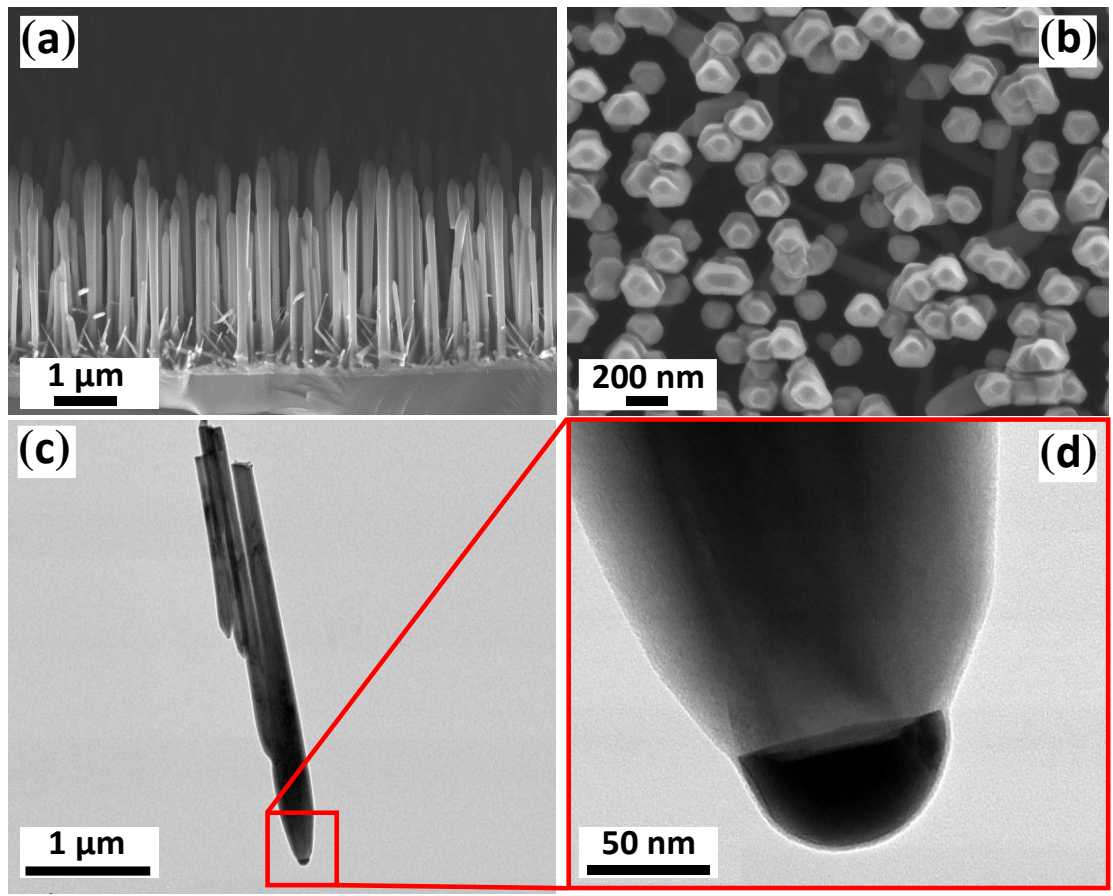

**FIG. 1.** Typical (a) cross-sectional and (b) planar FESEM images of the Ge-Ge$_{0.92}$Sn$_{0.08}$ core-shell NW arrays grown using a Ge-core. (c) A low-magnification TEM image of a few core-shell nanowires and (d) a magnified view of the tip of a single nanowire.

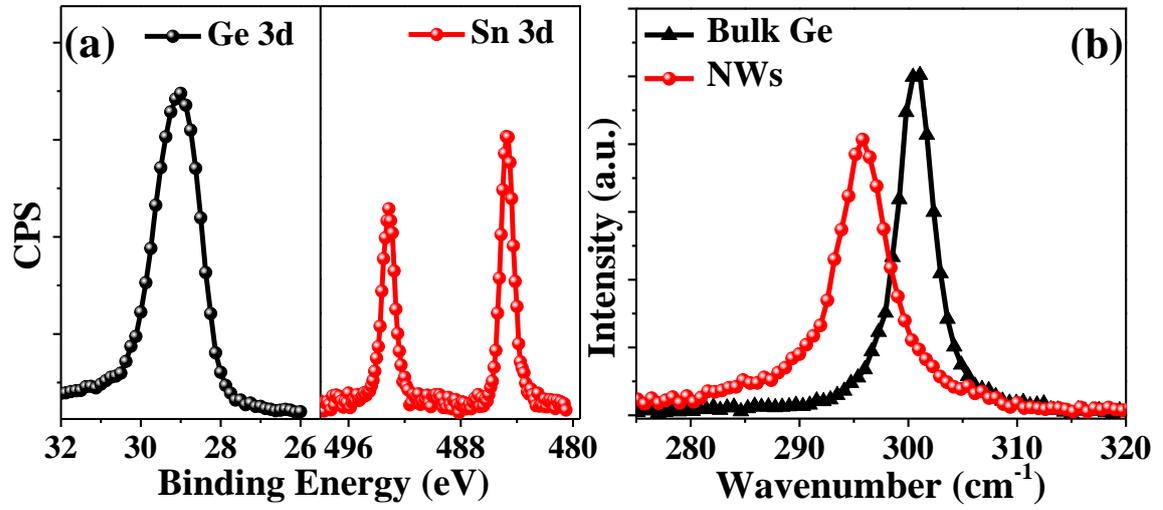

**FIG. 2.** (a) High-resolution X-ray photoelectron spectra of Ge-*3d* and Sn-*3d* core electrons of Ge-Ge$_{0.92}$Sn$_{0.08}$ core-shell NW arrays. (b) Room temperature Raman spectra of bulk Ge and Ge-Ge$_{0.92}$Sn$_{0.08}$ core-shell NWs.

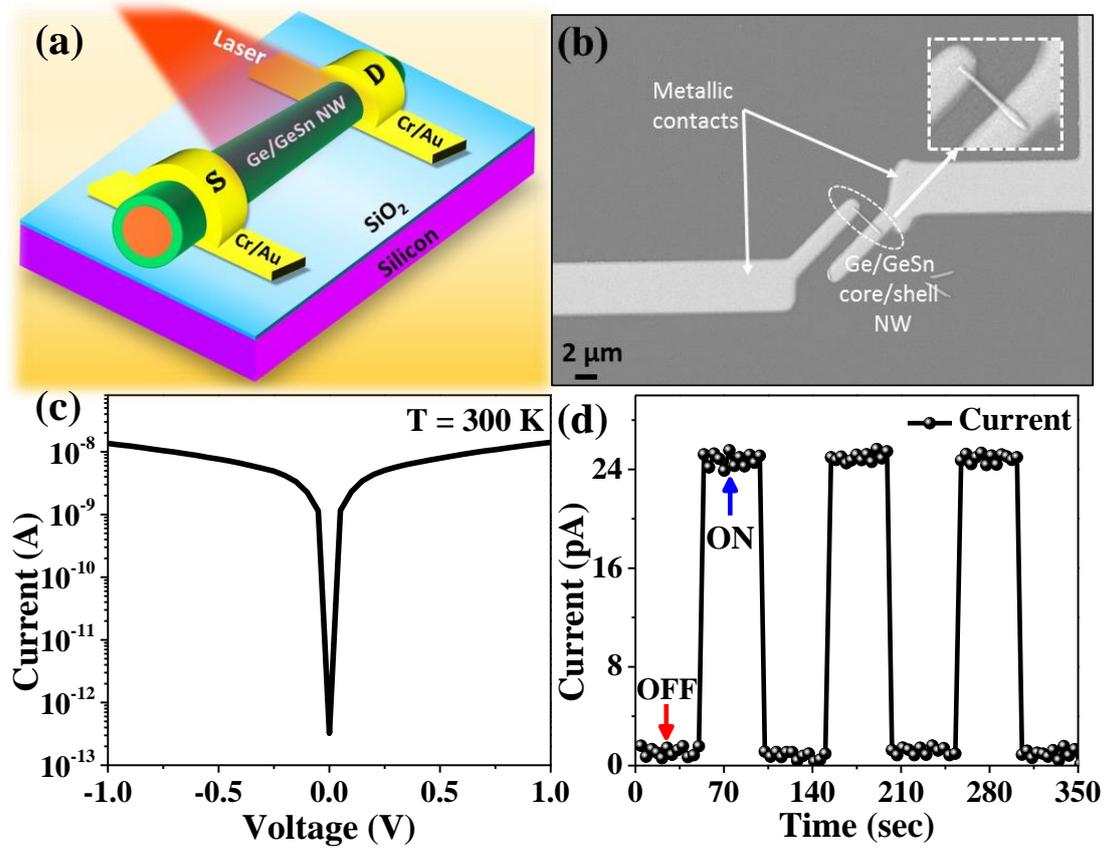

**FIG. 3.** (a) Schematic illustration and (b) corresponding FESEM micrograph of the fabricated Ge-Ge$_{0.92}$Sn$_{0.08}$ core-shell single NW photodetector device (inset showing a zoomed image). (c) Typical room-temperature semi-logarithmic current-voltage (*I-V*) curve of the fabricated single NW device measured under dark. (d) Temporal photoresponse of the single NW photodetector at a wavelength of 1.55 μm measured at zero bias under the excitation with an incident light intensity of ~6.37 mW/cm$^2$.

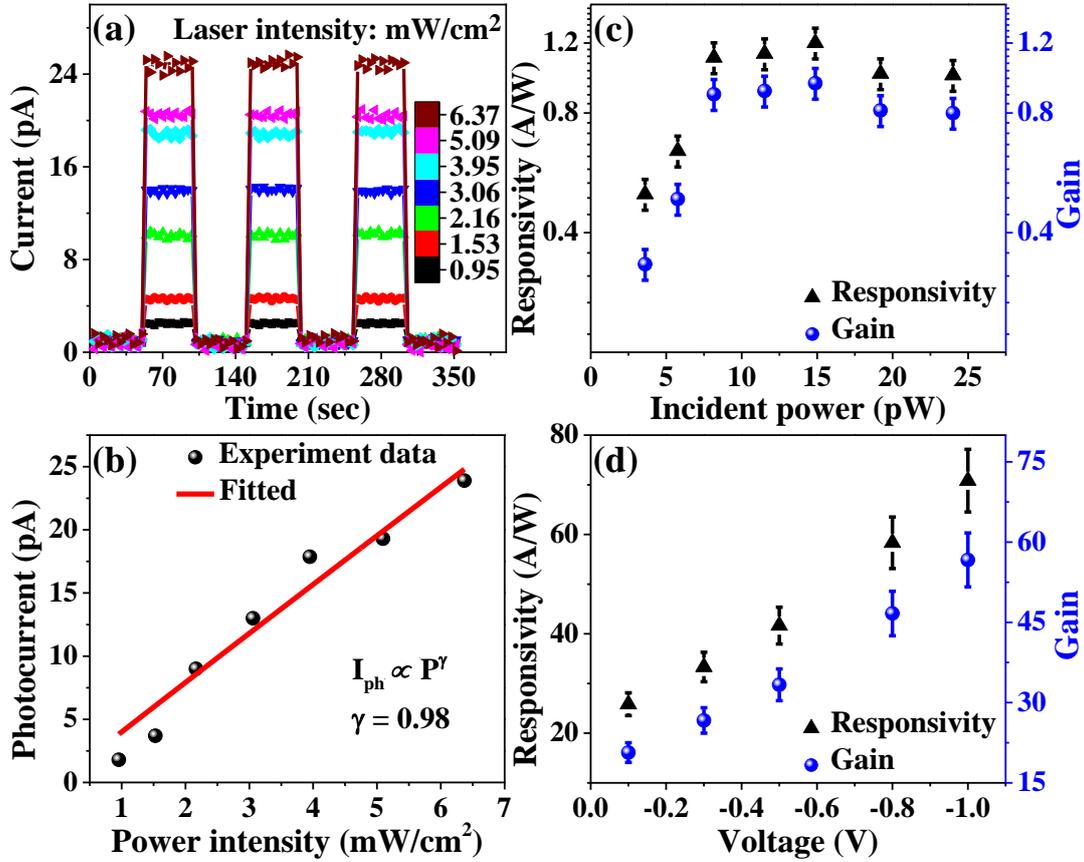

**FIG. 4.** (a) Photocurrent modulation of the Ge-Ge$_{0.92}$Sn$_{0.08}$ core-shell single NW photodetector as a function of excitation intensity measured at zero bias and (b) power law dependence of the photocurrent vs. incident power density. Responsivity and photoconductive gain at optical communication wavelength (λ – 1.55 μm) of the fabricated single NW photodetector (c) for different optical power at zero bias and (d) as a function of applied bias with constant illumination power of ~24 pW. Error bar represent the maximum estimated propagating error in each measurement.

# Supplementary material

# Ge-Ge$_{0.92}$Sn$_{0.08}$ core-shell single nanowire infrared photodetector with superior characteristics for on-chip optical communication

Sudarshan Singh[1], Subhrajit Mukherjee[1,2], Samik Mukherjee[3,¥], Simone Assali[3], Lu Luo[3], Samaresh Das[4], Oussama Moutanabbir[3], and Samit K Ray[1]

**AFFILIATIONS**

[1]Department of Physics, Indian Institute of Technology Kharagpur, Kharagpur, West Bengal 721302, India

[2]Presently at Faculty of Materials Science and Engineering, Technion-Israel Institute of Technology, Haifa, 3203003, Israel

[3]Department of Engineering Physics, École Polytechnique de Montréal, C. P. 6079, Succ. Centre-Ville, Montreal, Québec H3C 3A7, Canada

[¥]Present affiliation: Leibniz Institute for Solid State and Materials Research, Helmholtzstraße. 20, 01069 Dresden, Germany

[4]Centre for Applied Research in Electronics, Indian Institute of Technology Delhi, New Delhi 110016, India

Email: physkr@phy.iitkgp.ac.in

**FESEM images of Ge-Ge$_{0.92}$Sn$_{0.08}$ core-shell NWs and length histogram:**

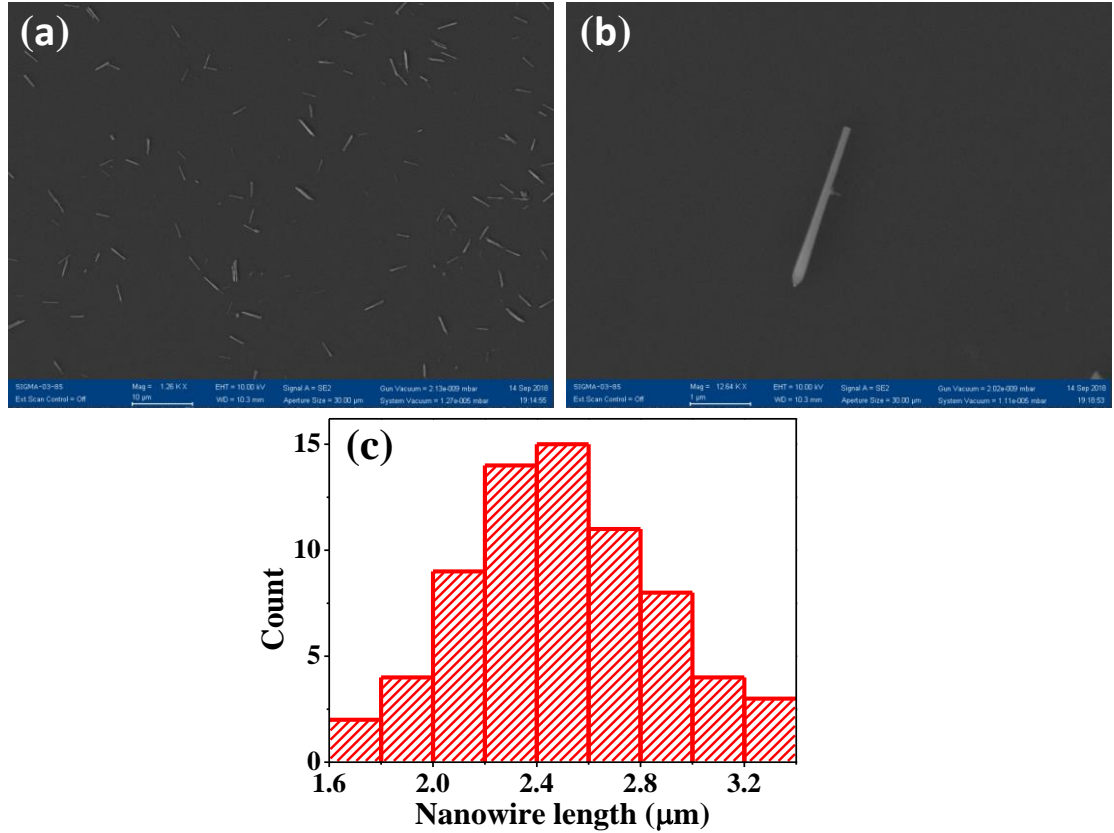

**FIG. S1.** (a) Field-emission scanning electron microscopic (FESEM) images of the Ge-Ge$_{0.92}$Sn$_{0.08}$ core-shell nanowires dispersed onto a SiO$_2$ wafer and (b) an isolated single nanowire. (c) Length histogram of the dispersed Ge-Ge$_{0.92}$Sn$_{0.08}$ core-shell nanowires revealing an average length in between 2-3 μm.

**Decoupling the Sn composition and strain in Ge-Ge$_{0.92}$Sn$_{0.08}$ core-shell nanowires using Raman spectrum:**

To decouple the Sn content and strain in the GeSn shell, the Raman spectrum of the Ge-Ge$_{0.92}$Sn$_{0.08}$ core-shell NWs has been fitted using two exponentially modified Gaussian (EMG) functions[1]. In addition to Ge-Ge LO phonon peak, a shoulder peak at lower wavenumber side is related to disorder-activated (DA) vibrational modes in GeSn shell[1]. Using the following equation describing the behaviour of the peak position ($\omega$) and asymmetry ($t$) of the main Ge-Ge mode,

$$\omega_{Ge-Ge} = \omega_{0,Ge-Ge} + a^{\omega}_{Ge-Ge}y + b^{\omega}_{Ge-Ge}\varepsilon \quad \ldots (S1)$$

$$t_{Ge-Ge} = t_{0,Ge-Ge} + a^{t}_{Ge-Ge}y + b^{t}_{Ge-Ge}\varepsilon \quad \ldots (S2)$$

The estimated value of Sn content ($y$) and strain ($\varepsilon$) are found to be around 8.3 % and -2.6 %, respectively.

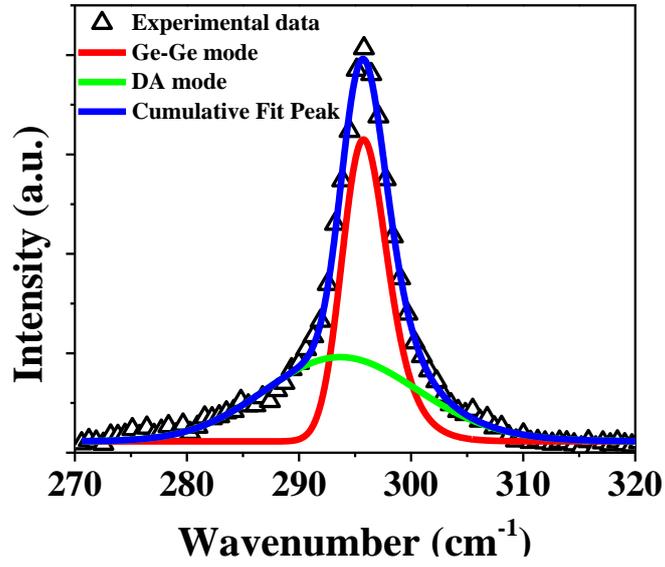

**FIG. S2.** Room-temperature Raman spectrum of Ge-Ge$_{0.92}$Sn$_{0.08}$ core-shell NWs.

**Noise current expression:**

The noise current has been estimated using the following relation[2]

$$\langle i_n^2 \rangle = \left[2qi_d + \frac{4k_BT}{R}\right]\Delta f \quad \ldots (S3)$$

$i_n$ = Noise current,
$i_d$ = Dark current of the detector,
$k_B$ = Boltzmann constant
$T$ = Temperature
$R$ = Shunt resistance
$\Delta f$ = Electrical Bandwidth (assuming 1Hz)

**TABLE SI: Performance comparison of Ge-Ge$_{0.92}$Sn$_{0.08}$ core-shell and various other single NW based photodetectors operating at a wavelength of 1.55 μm.**

| Material | Structure | Responsivity (A/W) @ 1.55 $\mu$m | Detectivity (Jones) | Photoconductive gain | Year | Ref |
|---|---|---|---|---|---|---|
| Si | NW array | 100 (@170 K) | - | - | 2010 | 3 |
| GaAsSb | NW | 1.44 | $6.55 \times 10^8$ | - | 2015 | 4 |
| InAs | NW | ~15 | ~ $6 \times 10^{11}$ | - | 2016 | 5 |
| InGaAs | Core/shell NW | 5.75 | - | - | 2017 | 6 |
| Ge | NW | 22.6 | | 2000 | 2017 | 7 |
| GaSb | NW | 61 | $7.6 \times 10^7$ | 49 | 2019 | 8 |
| GaAs$_{1-x}$Sb$_x$/InAs | Core/shell NW | 0.02 | - | - | 2019 | 9 |
| GeSn/Ge | Dual-NW | 0.0012 | - | - | 2020 | 10 |
| **Ge/GeSn** | **Core/shell NW** | **70.8** | **8.4×10$^9$** | **~57** | | **This work** |

**TABLE SII: Performance comparison of various GeSn-based photodetectors, fabricated with different Sn content, operating at a wavelength of 1.55 μm.**

| GeSn (Sn %) | Device structure | Responsivity (A/W) @ 1.55 $\mu$m | Detectivity (Jones) | Year | Ref |
|---|---|---|---|---|---|
| 2 % | PIN | 0.05 | - | 2009 | [11] |
| 0.5 % | PIN | 0.1 | - | 2011 | [12] |
| 2 % | PIN | 0.12 | - | 2011 | [13] |
| 3 % | PIN | 0.23 | - | 2011 | [14] |
| 4 % | PIN | 0.2 | - | 2012 | [15] |
| 9 % | QW/PC | 1.0 | - | 2012 | [16] |
| 3.9 % | PIN | 0.27 | - | 2013 | [17] |
| 8 % | MSM | 0.03 | - | 2013 | [18] |
| 1.8 % | PIN | 0.19 | - | 2014 | [19] |
| 4.2 % | PIN | 0.22 | - | 2014 | [20] |
| 7 % | QW PIN | 0.13 | - | 2014 | [21] |
| 7 % | PC | 0.18@10V | $1 \times 10^9$ | 2014 | [22] |
| 10 % | PC | 1.63@50V | $4.6 \times 10^9$ | 2014 | [23] |
| 5 % | PIN | 0.18 | - | 2015 | [24] |
| 10 % | PC | 2.85@5V | $4 \times 10^9$ | 2015 | [25] |
| 2.5 % | PIN | 0.45 | - | 2016 | [26] |
| 6 % | PIN | 0.5 | - | 2016 | [27] |
| 7 % | DHS PIN | 0.3 | $4 \times 10^9$ | 2016 | [28] |
| 10 % | DHS PIN | 0.19 | $2.4 \times 10^9$ | 2016 | [28] |
| 3 % | QW PIN | 0.080 | - | 2017 | [29] |
| 10 % | MWQ PIN | 0.216 | - | 2017 | [30] |
| 6.5 % | HPT | 1.8 | - | 2017 | [31] |
| 12.5 % | PC | 16.1 | $1.1 \times 10^{10}$ | 2019 | [32] |
| 6.5 % | PC | 140 | - | 2019 | [33] |
| 10 % | Side gated dual NW PD | 0.0012 | - | 2020 | [10] |
| **8 %** | **NW MSM** | **70.8@1V** | **8.4×10$^9$@1V** | | **This work** |

**References:**

[1] Bouthillier, S. Assali, J. Nicolas, and O. Moutanabbir, Semicond. Sci. Technol. **35**, (2020).

[2] H. Radamson and L. Thylén, in *Monolith. Nanoscale Photonics–Electronics Integr. Silicon Other Gr. IV Elem.* (Elsevier, 2015), pp. 63–85.

[3] A. Zhang, H. Kim, J. Cheng, and Y.H. Lo, Nano Lett. **10**, 2117 (2010).

[4] Z. Li, X. Yuan, L. Fu, K. Peng, F. Wang, X. Fu, P. Caroff, T.P. White, H. Hoe Tan, and C. Jagadish, Nanotechnology **26**, (2015).

[5] H. Fang, W. Hu, P. Wang, N. Guo, W. Luo, D. Zheng, F. Gong, M. Luo, H. Tian, X. Zhang, C. Luo, X. Wu, P. Chen, L. Liao, A. Pan, X. Chen, and W. Lu, Nano Lett. **16**, 6416 (2016).

[6] C. Zhou, X.T. Zhang, K. Zheng, P.P. Chen, W. Lu, and J. Zou, Nano Lett. **17**, 7824 (2017).

[7] U. Otuonye, H.W. Kim, and W.D. Lu, Appl. Phys. Lett. **110**, (2017).

[8] J. Sun, M. Peng, Y. Zhang, L. Zhang, R. Peng, C. Miao, D. Liu, M. Han, R. Feng, Y. Ma, Y. Dai, L. He, C. Shan, A. Pan, W. Hu, and Z.X. Yang, Nano Lett. **19**, 5920 (2019).

[9] X. Wang, D. Pan, Y. Han, M. Sun, J. Zhao, and Q. Chen, ACS Appl. Mater. Interfaces **11**, 38973 (2019).

[10] Y. Yang, X. Wang, C. Wang, Y. Song, M. Zhang, Z. Xue, S. Wang, Z. Zhu, G. Liu, P. Li, L. Dong, Y. Mei, P.K. Chu, W. Hu, J. Wang, and Z. Di, Nano Lett. **20**, 3872 (2020).

[11] J. Mathews, R. Roucka, J. Xie, S.Q. Yu, J. Meńdez, and J. Kouvetakis, Appl. Phys. Lett. **95**, 8 (2009).

[12] J. Werner, M. Oehme, M. Schmid, M. Kaschel, A. Schirmer, E. Kasper, and J. Schulze, Appl. Phys. Lett. **98**, 2009 (2011).

[13] R. Roucka, J. Mathews, C. Weng, R. Beeler, J. Tolle, J. Menéndez, and J. Kouvetakis, IEEE J. Quantum Electron. **47**, 213 (2011).

[14] S. Su, B. Cheng, C. Xue, W. Wang, Q. Cao, H. Xue, W. Hu, G. Zhang, Y. Zuo, and Q. Wang, Opt. Express **19**, 6400 (2011).

[15] M. Oehme, M. Schmid, M. Kaschel, M. Gollhofer, D. Widmann, E. Kasper, and J. Schulze, Appl. Phys. Lett. **101**, (2012).

[16] A. Gassenq, F. Gencarelli, J. Van Campenhout, Y. Shimura, R. Loo, G. Narcy, B. Vincent, and G. Roelkens, Opt. Express **20**, 27297 (2012).

[17] H.H. Tseng, H. Li, V. Mashanov, Y.J. Yang, H.H. Cheng, G.E. Chang, R.A. Soref, and G. Sun, Appl. Phys. Lett. **103**, (2013).

[18] J.Y.J. Lin, S. Gupta, Y.C. Huang, Y. Kim, M. Jin, E. Sanchez, R. Chen, K. Balram, D. Miller, J. Harris, and K. Saraswat, Dig. Tech. Pap. - Symp. VLSI Technol. 32 (2013).

[19] Y.H. Peng, H.H. Cheng, V.I. Mashanov, and G.E. Chang, Appl. Phys. Lett. **105**, (2014).

[20] M. Oehme, K. Kostecki, K. Ye, S. Bechler, K. Ulbricht, M. Schmid, M. Kaschel, M. Gollhofer, R. Körner, W. Zhang, E. Kasper, and J. Schulze, Opt. Express **22**, 839 (2014).

[21] M. Oehme, D. Widmann, K. Kostecki, P. Zaumseil, B. Schwartz, M. Gollhofer, R. Koerner, S. Bechler, M. Kittler, E. Kasper, and J. Schulze, Opt. Lett. **39**, 4711 (2014).

[22] B.R. Conley, A. Mosleh, S.A. Ghetmiri, W. Du, R.A. Soref, G. Sun, J. Margetis, J. Tolle, H.A. Naseem, and S.-Q. Yu, Opt. Express **22**, 15639 (2014).

[23] B.R. Conley, J. Margetis, W. Du, H. Tran, A. Mosleh, S.A. Ghetmiri, J. Tolle, G. Sun, R. Soref, B. Li, H.A. Naseem, and S.Q. Yu, Appl. Phys. Lett. **105**, (2014).

[24] Y. Dong, W. Wang, D. Lei, X. Gong, Q. Zhou, S.Y. Lee, W.K. Loke, S.-F. Yoon, E.S. Tok, G. Liang, and Y.-C. Yeo, Opt. Express **23**, 18611 (2015).

[25] T.N. Pham, W. Du, B.R. Conley, J. Margetis, G. Sun, R.A. Soref, J. Tolle, B. Li, and S.Q. Yu, Electron. Lett. **51**, 854 (2015).

[26] C. Chang, H. Li, S.H. Huang, H.H. Cheng, G. Sun, and R.A. Soref, Appl. Phys. Lett. **108**, 0 (2016).

[27] J. Zheng, S. Wang, Z. Liu, H. Cong, C. Xue, C. Li, Y. Zuo, B. Cheng, and Q. Wang, Appl. Phys. Lett. **108**, 1 (2016).

[28] T. Pham, W. Du, H. Tran, J. Margetis, J. Tolle, G. Sun, R.A. Soref, H.A. Naseem, B. Li, and S.-Q. Yu, Opt. Express **24**, 4519 (2016).

[29] M. Morea, C.E. Brendel, K. Zang, J. Suh, C.S. Fenrich, Y.C. Huang, H. Chung, Y. Huo, T.I.

Kamins, K.C. Saraswat, and J.S. Harris, Appl. Phys. Lett. **110**, 0 (2017).
[30] Y. Dong, W. Wang, S. Xu, D. Lei, X. Gong, X. Guo, H. Wang, S.-Y. Lee, W.-K. Loke, S.-F. Yoon, and Y.-C. Yeo, Opt. Express **25**, 15818 (2017).
[31] W. Wang, Y. Dong, S.-Y. Lee, W.-K. Loke, D. Lei, S.-F. Yoon, G. Liang, X. Gong, and Y.-C. Yeo, Opt. Express **25**, 18502 (2017).
[32] H. Tran, T. Pham, J. Margetis, Y. Zhou, W. Dou, P.C. Grant, J.M. Grant, S. Al-Kabi, G. Sun, R.A. Soref, J. Tolle, Y.H. Zhang, W. Du, B. Li, M. Mortazavi, and S.Q. Yu, ACS Photonics **6**, 2807 (2019).
[33] F. Yang, K. Yu, H. Cong, C. Xue, B. Cheng, N. Wang, L. Zhou, Z. Liu, and Q. Wang, ACS Photonics **6**, 1199 (2019).